\documentclass{emulateapj}   

\RequirePackage{natbib}
\usepackage{color}
\usepackage{xcolor}
\usepackage[figuresright]{rotating}
\usepackage{hyperref}
\usepackage{wrapfig} 
\usepackage{graphicx} 
\usepackage{epsfig}

\newcommand{\yt}{\texttt yt}
 
\newcommand{\Mach}{\mathcal{M}}

\usepackage{color}

\begin{document}


\title {Cosmological MHD Simulations of Galaxy Cluster Radio Relics: Insights
and Warnings for Observations}

\author{Samuel~W.~Skillman\altaffilmark{1,2}, Hao Xu\altaffilmark{3},
  Eric J.  Hallman\altaffilmark{1,4,5}, Brian W.
  O'Shea\altaffilmark{6,7}, Jack O. Burns\altaffilmark{1,4}, Hui
  Li\altaffilmark{3}, David C. Collins\altaffilmark{3}, Michael
  L. Norman\altaffilmark{8}}

\altaffiltext{1}{Center for Astrophysics and Space Astronomy, Department of Astrophysical \& Planetary Science, University of Colorado, Boulder, CO 80309}
\altaffiltext{2}{DOE Computational Science Graduate Fellow}
\altaffiltext{3}{Theoretical Division, Los Alamos National Laboratory, Los Alamos, NM 87544}
\altaffiltext{4}{Lunar University Network for Astrophysics Research (LUNAR), NASA Lunar Science Institute, NASA Ames Research Center, Moffet Field, CA, 94089} 
\altaffiltext{5}{Tech-X Corporation, Boulder, CO 80303} 
\altaffiltext{6}{Department of Physics \& Astronomy, Michigan State University, East Lansing, MI, 48824} 
\altaffiltext{7}{Lyman Briggs College and Institute for Cyber-Enabled Research, Michigan State University, East Lansing, MI, 48824} 
\altaffiltext{8}{Center for Astrophysics and Space Sciences, University of California at San Diego, La Jolla, CA 92093, USA}

\email{samuel.skillman@colorado.edu}

\begin{abstract}

Non-thermal radio emission from cosmic ray electrons in the vicinity
of merging galaxy clusters is an important tracer of cluster merger
activity, and is the result of complex physical processes that
involve magnetic fields, particle acceleration, gas dynamics, and
radiation.  In particular, objects known as radio relics are thought
to be the result of shock-accelerated electrons that, when embedded
in a magnetic field, emit synchrotron radiation in the radio
wavelengths.  In order to properly model this emission, we utilize the adaptive mesh refinement
simulation of the magnetohydrodynamic evolution of a galaxy cluster
from cosmological initial conditions.  We locate shock fronts and apply
models of cosmic ray electron acceleration that are then input into
radio emission models.  We have determined
the thermodynamic properties of this radio-emitting plasma and
constructed synthetic radio observations to compare to observed
galaxy clusters.  We find a significant dependence of the observed
morphology and radio relic properties on the viewing angle of the
cluster, raising concerns regarding the interpretation of observed
radio features in clusters.  We also find that a given shock should
not be characterized by a single Mach number.  We find that the bulk 
of the radio emission comes from gas
with $\mathrm{T}>5\times10^7\mathrm{K}$, $\mathrm{\rho} \sim
10^{-28}-10^{-27} \mathrm{g/cm^3}$, with magnetic field strengths of
$0.1-1.0 \mathrm{\mu G}$ and shock Mach numbers of $\Mach \sim
3-6$.  We present an analysis of the radio spectral index which
suggests that the spatial variation of the spectral index can mimic
synchrotron aging.  Finally, we examine the polarization
fraction and position angle of the simulated radio features, and compare to
observations.

\end{abstract}


\section{Introduction}

Galaxy clusters are hosts to a variety of thermal and non-thermal phenomena,
many of which are the result of cosmological structure formation.  The study of
relativistic particles in galaxy cluster environments was motivated by the
observation of the radio halo in the Coma cluster by
\citet{1959Natur.183.1663L}, and has since grown into an industry of
observations, theory, and simulation. For a review on current radio
observations of galaxy clusters see \citet{2008SSRv..134...93F, 2012A&ARv..20...54F}, and for
a review on the non-thermal processes see \citet{2008SSRv..134..311D}.  Here we
review the basic characteristics of galaxy cluster radio ``halos'' and giant
radio ``relics,'' to use the classification in \citet{2008SSRv..134...93F}.
Radio halos are usually $\sim\mathrm{Mpc-}$sized features in galaxy clusters,
closely following the X-ray morphology in the central regions of the cluster.
They are generally characterized by very low ($<$ few percent) linear polarization
fractions, and are found in galaxy clusters with disturbed morphology and no
evidence for a cool core \citep{2008SSRv..134...93F, 2009A&A...507.1257G,
2012A&ARv..20...54F}.  The
origin of the emission is thought to be from relativistic ($\gamma\sim10^4$)
electrons emitting synchrotron radiation.  The source of the energy in these
electrons, however, is debated.  It may originate from the decay of pions (the
``secondary'' or ``hadronic'' model), created by interactions between cosmic
ray protons and the thermal population \citep{1980ApJ...239L..93D,
2000A&A...362..151D, 2001ApJ...562..233M}, which would be strengthened by the
observation of gamma-ray emission in cluster cores.  However, initial studies 
of many galaxy clusters using the FERMI satellite \citep{2010ApJ...717L..71A}, 
as well as for fewer objects with other instruments \citep[e.g. MAGIC observations
of Perseus][]{2010ApJ...710..634A,2012A&A...541A..99A}, combined with radio data \citep{2011ApJ...728...53J, 2012MNRAS.426..956B}
constrain the energy in cosmic rays to be very low $(\ll 10\%)$ of the thermal
energy in most cases.  Others believe that the electrons are turbulently
accelerated either from the thermal population or from aging populations of
electrons either from shock acceleration or AGN/supernova injection
\citep{2011MNRAS.410..127B}.

Radio relics, on the other hand, are thought to be accelerated by first-order
Fermi acceleration through the process of Diffusive Shock Acceleration (DSA)
\citep{1978ApJ...221L..29B}.  They have a relatively steep radio, and therefore
inferred electron, spectrum where $S\propto \nu^{-\alpha}$ with $\alpha \approx
1-2$.  These radio sources are not associated with any of the cluster galaxies
or AGN bubbles.  They are also not associated with any point sources in other
wavelengths, and are usually found in the outskirts of clusters.  Their
location can be up to $\sim 2~\mathrm{Mpc}$ from the cluster core, and can be
extended up to $\sim1.5~\mathrm{Mpc}$ in length \citep{2010Sci...330..347V,
2011A&A...528A..38V, 2012MNRAS.425L..36V}.  In some cases these radio features
are coincident with X-ray surface brightness and temperature jumps, potentially
indicating the presence of a shock front \citep{2010ApJ...715.1143F, 2012PASJ...64...67A}.  While
more rare, double radio relics are observed in several systems.  Double radio
relics are unique in that they provide tighter constraints on the geometry and
kinematics of the merging clusters\citep{2011MNRAS.418..230V}.  Upcoming radio
telescopes such as LOFAR, the Jansky VLA, and eventually the SKA will provide
an increase in sensitivity and resolution (both spectral and spatial) that will
allow for discoveries in blind surveys.  Because of this, we are at an
important time to use simulation and theory to predict the number and the
properties of relics in cosmological samples.  Past simulations have focused on
both single clusters \citep{1999ApJ...518..603R, 2008MNRAS.385.1211P,
2009MNRAS.393.1073B} as well as ensembles of clusters
\citep{2008MNRAS.391.1511H, 2011ApJ...735...96S, 2012MNRAS.421.1868V, 2012MNRAS.420.2006N}.  Both
are needed in order to constrain the plasma physics and how varying
environments lead to observational quantities such as luminosity functions.  

In this paper we investigate the origins, properties, and observational
implications of a merging galaxy cluster using a numerical simulation.  For the
first time, we start from cosmological initial conditions and self-consistently
evolve the cluster magnetic field from an AGN source the equations of
magnetohydrodynamics rather than assuming a magnetic field strength and
topology.  This allows us to explore one scenario in which the magnetic field
forms, evolves, and interacts with the radio relic emission.  We describe, in
detail, the plasma environment of the radio-emitting regions.  

After investigating the properties of the cluster gas, we analyzed the
resulting radio emission using novel approaches to explore systematic effects
present in current radio observations.  We used a new tool to view this
Eulerian grid simulation from arbitrary directions in order to demonstrate the
effect of viewing angle on the derived properties.  We then developed the
capability to integrate the polarized radio emission along the line of sight to
provide the closest comparison to observations. We then use this technique to
produce polarization fraction and position angle maps from our MHD AMR
simulation, and provide comments on the relevance of our results to current
observations of radio features in galaxy clusters.  Finally, we discuss the
impact of using previous assumptions about the magnetic field compared to the
values that are self-consistently evolved from an AGN source.  We use this to
provide insight into observational results.

\section{Methods}\label{sec:methods}

\subsection{Simulations}\label{sec:simulations}

Our simulation was run using a modified version of the \textit{Enzo} cosmology
code \citep{1997astro.ph.10187B, 1997ASPC..123..363B, 1999ASSL..240...19N,
2004astro.ph..3044O}.  \textit{Enzo} uses block-structured adaptive mesh
refinement \citep[AMR;][]{1989JCoPh..82...64B} as a base upon which it couples
an Eulerian hydrodynamic solver for the gas with an N-Body particle mesh (PM)
solver \citep{1985ApJS...57..241E, 1988csup.book.....H} for the dark matter.
In this work we utilize the MHD solver described in
\citet{2010ApJS..186..308C}. The solver employed here is spatially second
order, while the PPM solver \citep{1984JCoPh..54..174C} commonly used in Enzo is spatially
3rd order.  The net effect on the shock-finding algorithm will be to broaden
a single shock by a small amount.  However it will be impossible to disentangle
this effect from the changes in shock structure due to the addition of magnetic
forces in the evolution.  None of the results we present here will be sensitive
to these small differences. We have extended this version of \textit{Enzo} to
include temperature-jump based shock-finding as described in
\citet{2008ApJ...689.1063S} and used in \citet{2011ApJ...735...96S}. 

The galaxy cluster studied in this work is the same as cluster U1 in
\citet{2011ApJ...739...77X}. In this work, clusters were formed from
cosmological initial conditions, and magnetic fields were injected by the most
massive galaxy at a variety of stages in the cluster evolution. It was found
that different injection parameters of magnetic fields have little impact on
the cluster formation history.  This simulation models the evolution of dark
matter, baryonic matter, and magnetic fields self-consistently.  The simulation
uses an adiabatic equation of state for gas, with the ratio of specific heat
being 5/3, and does not include heating or cooling physics or chemical
reactions.  While studies have been done including these physical models and
their role in characterizing shocks \citep{2007ApJ...669..729K,
2007MNRAS.378..385P}, we chose to ignore them due to both computational cost as
well as possible confusion between structure formation shocks and those arising
from star/galaxy feedback. Additionally, \citet{2007ApJ...669..729K} found
little effect on the overall kinetic energy dissipation between simulations
with adiabatic gas physics and those including cooling and feedback, and while
\citet{2007MNRAS.378..385P} show changes at high Mach number, as we will see
these have little consequence for the shocks involved with producing radio
relics.

The initial conditions of the simulation
are generated at redshift $z=30$ from an \citet{1999ApJ...511....5E} power
spectrum of density fluctuations in a $\Lambda$CDM universe with parameters
$h=0.73$, $\Omega_{m}=0.27$, $\Omega_{b}=0.044$, $\Omega_{\Lambda}=0.73$,
$\sigma_{8}=0.77$, and $n_{s}=0.96$. These parameters are close to the values
from WMAP3 observations \citep{2003ApJS..148..175S}.  While these parameters
differ from the latest constraints, it is largely irrelevant for this
particular project. The simulated volume is ($256$ $h^{-1}$Mpc)$^{3}$, and it
uses a $128^3$ root grid and $2$ nested static grids in the Lagrangian region
where the cluster forms. This gives an effective root grid resolution of
$512^3$ cells ($\sim$ 0.69 Mpc) and dark matter particle mass resolution of
$1.07 \times 10^{10}M_{\odot}$.  During the course of the simulation, $8$
levels of refinements are allowed beyond the root grid, for a maximum spatial
resolution of $7.8125$ $h^{-1}$ kpc. The AMR is applied only in a region of
($\sim$ 43 Mpc)$^3$ where the galaxy cluster forms near the center of the
simulation domain. The AMR criteria in this simulation are the same as in
\citet{2011ApJ...739...77X}. During the cluster formation but before the
magnetic fields are injected, the refinement is only controlled by baryon and
dark matter density, refining on overdensities of 8 for each additional level.
After magnetic field injections, in addition to the density refinement, all the
regions where magnetic field strengths are higher than 5 $\times$ 10$^{-8}
\mathrm{G}$ are refined to the highest level.  The importance of using this
magnetic field refinement criterion in cluster MHD simulations is discussed in
\citet{2010ApJ...725.2152X}. 
 
The magnetic field initialization used is the same method in
\citet{2008ApJ...681L..61X, 2009ApJ...698L..14X} as the original magnetic tower
model proposed by \citet{2006ApJ...643...92L}, and assumes the magnetic fields
are from the outburst of AGN. The magnetic fields are injected at redshift
$z=3$ in two proto-clusters, which belong to two sub-clusters. The injection
locations are the same locations in simulations U1a and U1b in
\citet{2011ApJ...739...77X}.  There is $\sim$ 6 $\times$ 10$^{59}$ erg of
magnetic energy placed into the ICM from each injection, assuming that $\sim$
1 percent of the AGN outburst energy of a several 10$^8$ M$_{\odot}$ SMBH is in
magnetic fields. Previous studies \citep{2010ApJ...725.2152X} have shown that
the injection redshifts and magnetic energy have limited impact on the
distributions of the ICM magnetic fields at low redshifts. 

The simulated cluster is a massive cluster with its basic properties at
redshift $z=0$ as follows:  R$_{virial}$ = 2.5 Mpc, M$_{virial}$(total) = 1.9
$\times$ 10$^{15}$ M$_{\odot}$, M$_{virial}$(gas) = 2.7 $\times$ 10$^{14}$
M$_{\odot}$, and T$_{virial}$ = 10.3 keV. This cluster is in an unrelaxed
dynamical state at $z=0$ with its two magnetized sub-clusters of similar size
undergoing a merger. The total magnetic energy in the simulation at $z=0$ is
9.6 $\times$ 10$^{60}$ erg, nearly all of which is within the cluster virial
radius. The details about the cluster formation are described in
\citet{2011ApJ...739...77X}.

\subsection{Synchrotron Emission}\label{sec:synchrotron_emission} 

We use the same technique as was presented in \citet{2011ApJ...735...96S} and
based on \citet{2007MNRAS.375...77H}, except we no longer rely on the
assumption that the magnetic field is a simple function of density and instead
use the magnetic field from the simulation. This method assumes that a fraction
of the incoming kinetic energy of the gas is accelerated by the shock up to
a power-law distribution in energy, which extends from the thermal
distribution.  This distribution is that predicted by diffusive shock
acceleration theory in the test-particle limit \citep{1983RPPh...46..973D,1978MNRAS.182..147B, 2001RPPh...64..429M,2007A&A...475....1A,
1978ApJ...221L..29B, 2012arXiv1203.3681O}.  At the high energy end, it also
assumes that there is an exponential cutoff determined by the balance of
acceleration and cooling.  The total radio power from a shock wave of area $A$,
frequency $\nu_{obs}$, magnetic field $B$, electron acceleration efficiency
$\xi_e$, electron power-law index $s$ ($n_e \propto E^{-s}$), post-shock
electron density $n_e$ and temperature $T_2$ is \citep{2007MNRAS.375...77H} 
\begin{eqnarray} \label{eq:2}
\frac{dP(\nu_{obs})}{d\nu} = 6.4\times10^{34}erg\ s^{-1}\ Hz^{-1}
\frac{A}{\mathrm{Mpc}^2} \frac{n_e}{10^{-4} \mathrm{cm}^{-3}}\cr
\frac{\xi_e}{0.05}(\frac{\nu_{obs}}{1.4GHz})^{-s/2}  \times (\frac{T_2}{7
keV})^{3/2}\cr \frac{(B/\mu G)^{1+(s/2)}}{ (B_{CMB}/\mu G)^2 + (B/\mu G)^2}
\Psi (\Mach).  \end{eqnarray}
where $\Psi(\Mach)$ is a dimensionless shape function that rises steeply above
$\Mach\sim2.5$ and plateaus to $1$ above $\Mach \sim 10$. In all work presented
we use a fiducial value of $\xi_e=0.005$, as suggested in
\citet{2008MNRAS.391.1511H}, and the same as used in
\citet{2011ApJ...735...96S}. 

There are several important things to notice about this model, which will help
guide our interpretations of the results throughout this paper. First,
the emission scales linearly with the downstream electron density, and with the
downstream temperature to the $3/2$ power. Additionally, in regions where the
magnetic field is less than the equivalent magnetic field strength from the CMB
energy density, $B_{CMB}$, the emission scales with $B^{1+s/2}$,
where $s\sim3$ for most relic situations. 

\subsection{Analysis Tools}

In this work we relied heavily on the data analysis and visualization toolkit,
\textit{yt} \citep{2011ApJS..192....9T}, to produce the derived data products
presented.  Here we describe the tools used specifically in our analysis, and
leave further description to the \textit{yt} documentation
\footnote{http://yt-project.org/doc}. Derived
Quantities\footnote{http://yt-project.org/doc/analyzing/objects.html\#derived-quantities},
such as WeightedAverageQuantity and TotalQuantity, are used to calculate
weighted averages and totals of fluid quantities.  For example, we use
WeightedAverageQuantity to calculate the average temperature, weighted by cell
mass.  To analyze properties such as the radio and X-ray emission from our
simulations, we use Derived Fields
\footnote{http://yt-project.org/doc/analyzing/creating\_derived\_fields.html}
to define the functional form of our new fields, which is then calculated on
a grid-by-grid basis as needed.  To calculate distribution functions, either as
a function of position or fluid quantity, we utilize 1-D Profiles and 2-D Phase
Plots\footnote{http://yt-project.org/doc/visualizing/plots.html\#d-profiles}.
By specifying a binning field, we are then able to calculate either the total
of another quantity or the average (along with the standard deviation).  These
are used to create radial profiles as well as characterize quantities such as
the average magnetic field strength as a function of density and temperature.
We also take advantage of adaptive slices and projections.
Slices\footnote{http://yt-project.org/doc/visualizing/plots.html\#slices}
sample the data at the highest resolution data available, and return an
adaptive 2D image that can then be re-sampled into fixed resolution images.
Similarly, we use weighted and unweighted
projections\footnote{http://yt-project.org/doc/visualizing/plots.html\#projections}
of quantities to provide average or total quantities integrated along the line
of sight.  Again, these adaptive 2D data objects can then be re-sampled to
create images at various resolutions.  We also utilize and extend off-axis
projections for use in integrating the polarization vectors of radio emission,
to be described further in Section \ref{sect:polarization}.  Finally, we use
the spectral frequency integrator to calculate the X-ray emission based on the
Cloudy code, as was done in \citet{2011ApJ...735...96S} and \citet{
2011MNRAS.418.2467H}, and was described in detail in
\citet{2008MNRAS.385.1443S}.

\subsection{Polarization}\label{sect:polarization}

In addition to calculating the synchrotron emission, in this paper we
investigate the polarization fraction and position angles of the emission.  In
order to compare our simulations to observations, we have developed several new
tools, including the ability to calculate the polarization properties of the
radio emission. In this section, we describe how we calculate Stokes I, Q, and
U parameters from any viewing angle of our simulation.  For a review of these
topics, see \citet{1966MNRAS.133...67B, 1994hea..book.....L,
2002ASPC..278..131H}. While previous analyses of polarized emission were
capable of viewing along the coordinate directions, to our knowledge,
\textbf{this is the first presentation of off-axis polarized radio emission
from AMR simulations}\footnote{\citet{2008MNRAS.391.1511H} studied the
view dependence on the total radio emission}. This capability presents several
challenges.  Whereas the total radio emission is calculated as a direct sum of
the emission multiplied with the path length, the calculation of the polarized
emission requires simultaneous integration of each polarized component along
the line of sight  due to their mixing through Faraday rotation.

We have built this capability on top of the analysis package \textit{yt}.  We
began with the ``off-axis
projection"\footnote{http://yt-project.org/doc/cookbook/index.html\#cookbook-offaxis-projection}
operation, which is an off-axis ray-casting mechanism.  It operates by creating
a fixed-resolution image plane for which each pixel is then integrated through
the simulation volume. To do this correctly, first the AMR hierarchy is
homogenized into single-resolution bricks that uniquely tile the domain.  This
ensures that only the highest resolution data is used for a given point in
space.  These bricks are ordered and traversed by the image plane.  The result
of this is that we are able to integrate along the line of sight through the
AMR hierarchy sampling only the highest-resolution cells for that given point
in space.

We have furthermore modified this framework such that the RGB channels of the
image act as the total emission, I, and polarized emission along the x, $I_x$,
and y, $I_y$, axes. $I_x$ and $I_y$ can be thought of as the emission-weighted electric
field. The details of this calculation can be found in Appendix A. We first create derived
fields that correspond to the magnetic field projected onto the unit vectors
$\vec{v_x}$, $\vec{v_y}$ and $\vec{v_{||}}$, where $\vec{v_x}$ and $\vec{v_y}$
are defined with respect to east and north vectors defined by the viewing
direction, $\vec{v_{||}}$.  We label these magnetic fields as $B_x$, $B_y$, and $B_{||}$, 
respectively. We then define the polarization angle $\chi$ of the electric field as the
angle made between $B_x$ and $B_y$ rotated by $\pi/2$.  Finally, we define
a Faraday rotation field $\Delta \phi = 2.62\times10^{-17} \times \lambda^2 n_e B_{||}
\mathrm{dl}$, where all variables are in \texttt{cgs} units. Using a similar
notation to \cite{2009ApJ...693....1O}, we then integrate
along the line of sight the I, $I_x$, and $I_y$ values using the following discrete
step: 
\begin{eqnarray} 
  \label{eq:iqu} 
  \left[ { \begin{array}{cc} I_{i+1}  \\ I_{x, i+1} \\ I_{y, i+1}  \\ \end{array} } \right] 
  = \left[ 
    {\begin{array}{ccc} 
        dl &\ 0 &\ 0 \\ 
        dl\ f_p\ (\vec{v_x} \cdot \vec{E}) &\ cos(\Delta \phi) &\ -sin(\Delta \phi) \\ 
        dl\ f_p\ (\vec{v_y} \cdot \vec{E}) &\ sin(\Delta \phi) &\ cos(\Delta \phi)  \\ 
    \end{array} } 
  \right] 
  \left[ {\begin{array}{cc} \epsilon_{i}  \\ I_{x, i} \\ I_{y, i} \\ \end{array} } \right] 
\end{eqnarray} 
where $\Delta \phi$ is the Faraday rotation, $\vec{v_x}$ and $\vec{v_y}$ are
the image plane coordinate vectors, $f_p$ is the fractional polarization of the
synchrotron radiation of a given power-law slope of electrons, $dl$ is the ray
segment length between the incoming and outgoing face of each cell, and
$\vec{B}$ is the magnetic field.
Once integrated through the volume, we are then able to
create intensity, polarization fraction, and polarization direction maps.  This
capability is available to download using the changeset with hash fc3acb747162
 here: \url{https://bitbucket.org/samskillman/yt-stokes}.

\begin{figure*}[htp] \centering
\includegraphics[width=1.0\textwidth]{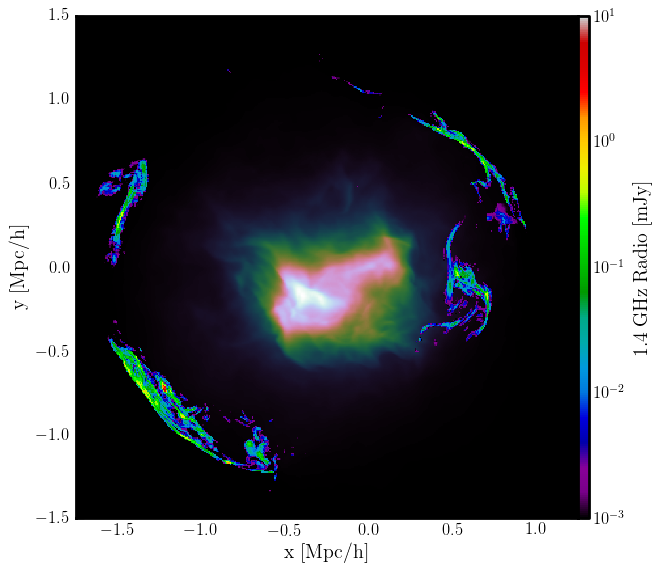}
\caption{Simulated X-ray and radio emission.  The X-ray in the central regions
shows a dynamic range of 100. The radio emission is calculated by placing the
simulated cluster at $100Mpc/h$, and masked to show
a dynamic range of $10^4$.} \label{fig:radio-xray-overlay}
\end{figure*}

\section{Radio Relic Properties}

In this section we describe the general properties of the simulated galaxy
cluster.  We begin by comparing the morphological similarities between our
simulated cluster and several observed clusters.  We then move on to describe
the other gas properties in an effort to constrain the properties of the
radio-emitting plasma.  Finally, we will look at the time evolution of these
quantities in order to understand their coupling during the merger process.

\subsection{Simulated Radio \& X-ray}

We begin by comparing the radio and X-ray emission from our simulation with the
radio relics present in A3376 (see Figure 1a. in \citet{2006Sci...314..791B})
and CIZA J2242.8+5301 (see Figure 1 in \citet{2010Sci...330..347V}). In this
work we calculate the X-ray emission using Cloudy to integrate the emission
from $0.5-12~\mathrm{keV}$ assuming a metallicity of $Z/Z_{solar}=0.3$.  The
resulting $1.4~\mathrm{GHz}$ radio and $0.5-12~\mathrm{keV}$ X-ray emission is
overlaid in Figure \ref{fig:radio-xray-overlay}.  The X-ray is shown in color
with a dynamic range of $100$.  The radio flux is calculated by placing the 
simulated cluster at a distance of $100 Mpc/h$. We then mask the radio emission such that
$10^{-3}-10^{1}~\mathrm{mJy}$~is visible.  The total integrated flux for the
left and right relics are $9.67\times10^{24}$ and $3.11\times10^{24} W/Hz$
at $1.4 GHz$, which is similar to many of the observed single and double radio
relics \citep{2012A&ARv..20...54F}.

There are a few specific details that we highlight here due to their
similarities to many observed radio relics.  First, we note that this appears
as a double radio relic.  These are relatively rare compared to their
single-sided counterparts.  This snapshot is following a major merger roughly
$300~\mathrm{Myr}$ after core passage.  The primary cluster is moving to the
lower-left, with the secondary moving primarily to the right.  After core
passage, a merger shock develops, moving both to the lower left and upper
right, and as will be seen in later figures, aligns with strong jumps in
temperature and density. The two relic features are aligned with the direction
of the merger as well as the shape of the X-ray emission.  This alignment of
the X-ray morphology and radio emission is characteristic of double radio
relics (see \citet{2011ApJ...735...96S}), as well as many single relics
\citep[e.g.][]{2011A&A...533A..35V, 2012MNRAS.426...40B}. 

The second key feature to this simulated double relic is the apparent aspect
ratio of the radio emission.  The length of the left relic (if we connect the
two pieces that are separated by a short distance) is more than
$2~\mathrm{Mpc/h}$, while the right-moving relic is $1.5-2.0~\mathrm{Mpc/h}$.
While these relics are quite elongated, they are very thin.  In many regions it
is at most $100-200~\mathrm{kpc/h}$ wide, even in projection.  We note here,
however, that this width is most likely underestimated since we are not
tracking the aged populations of electrons that would exist for some amount of
time behind the shock front.  

Comparing to A3376 \citep{2006Sci...314..791B}, we see many striking
resemblances, including the complex morphology of the Eastern portion of the
relic as well as the ring-like structure to the outline of the radio emission.
This, at the very least, suggests that the radio-emitting electrons are indeed
related to the shock structures formed in merging galaxy clusters.  We see
a similar structure in CIZA J2242.8+5301 \citep{2010Sci...330..347V}, where the
elongation of the X-ray emission points in the direction of the merger,
aligning with the double radio relic.  We also note the resemblance here with
our simulation in terms of the very thin region of radio emission along the
relic.  This suggests that the cooling times of the relativistic electrons must
be short.

\begin{sidewaysfigure*}[htp] \centering
\vspace{100 mm}
  \includegraphics[width=1.0\textwidth]{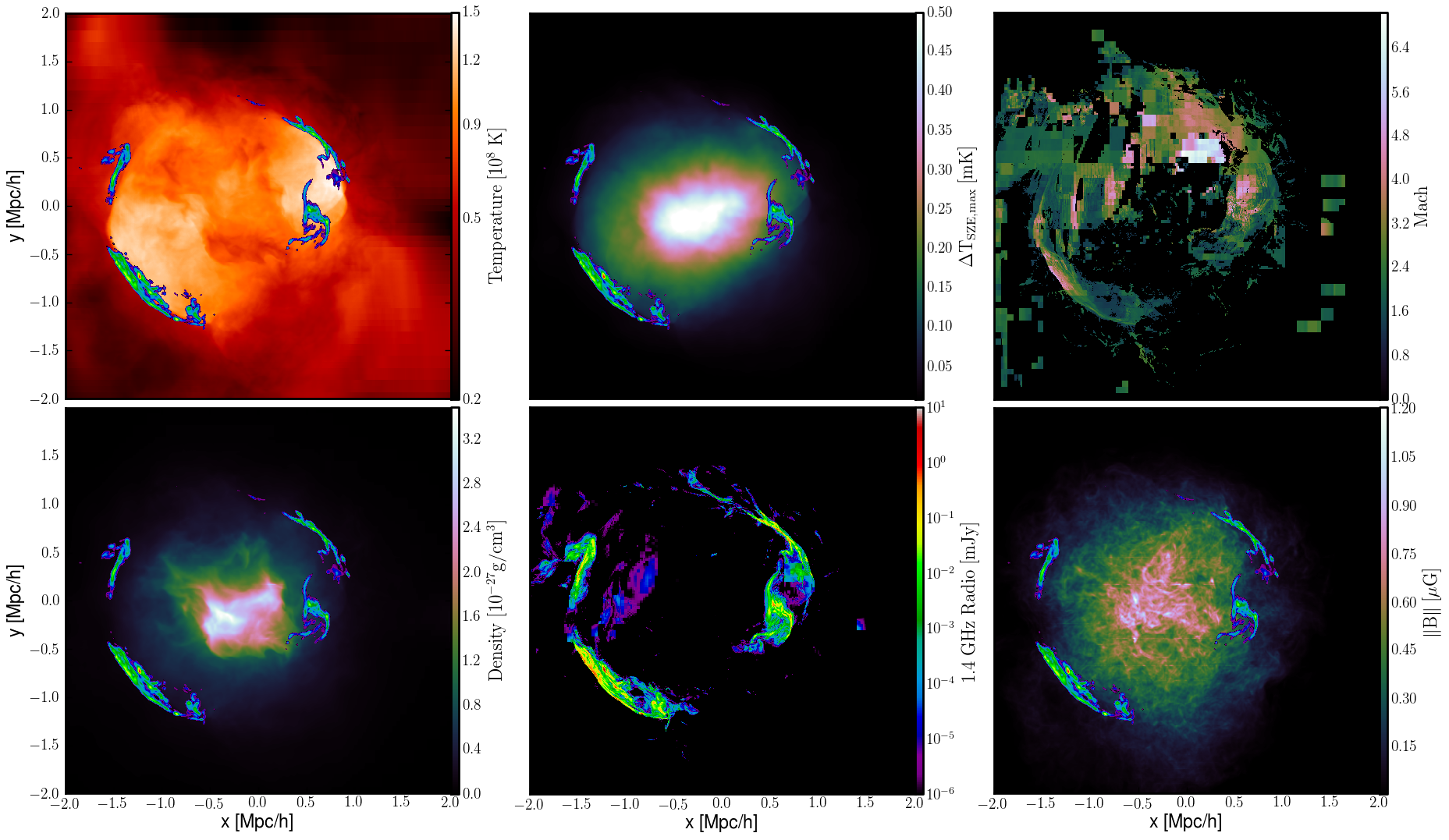} 
\caption{Projections of gas quantities.  Upper left: Temperature, weighted by
    density.  Upper middle: Sunyaev-Zel'dovich Compton-y value converted to the
    maximum temperature decrement.  Upper right: Mach number,weighted by radio
    emission. Lower left: Density, weighted by density. Lower middle:
    Integrated radio emissivity. Lower right: Magnetic field strength, weighted
    by density. Radio emission is overplotted on all panels except the Mach
    number, and is masked to show between $10^{-3}-10~\mathrm{mJy}$.  Radio
emission assumes cluster is at a distance of $100~\mathrm{Mpc/h}$.}
\label{fig:montage} 
\end{sidewaysfigure*}

Figure \ref{fig:montage} shows the fundamental quantities such as density,
temperature, Mach number, and magnetic field strength, along with observable
quantities such as the 1.4 GHz radio and temperature fluctuations in the CMB
due to the thermal Sunyaev-Zel'dovich effect.  The radio flux assumes
a distance to the simulated cluster of $100~\mathrm{Mpc/h}$. Each panel shows
the same field-of-view and depth of $4.0~\mathrm{Mpc/h}$ at z=0. In all panels
except for the radio and Mach number maps, we have overplotted the radio
emission to help guide the reader's eye in determining the location of the
emission relative to the underlying plasma.  The top left panel shows the
density-weighted temperature.  Here the correlation between the radio emission
and the temperature structure is very strong, as the outward moving shocks are
heating the gas to several $\times 10^8\mathrm{K}$.  Note that the sharp edges
in the temperature structure, as well as the maximum in the temperature
distribution, occurs $1.5-2\mathrm{Mpc/h}$ away from the center of the cluster.
This highlights the unrelaxed nature of this merging cluster.

The bottom-left panel shows the density-weighted density, and has a similar
structure to the X-ray image shown in Figure \ref{fig:radio-xray-overlay}, as
is expected since the X-ray emission is a strong function of gas density.  We
note that unlike the temperature, the density is strongly peaked towards the
center of the cluster, though the unrelaxed nature is evident by the elongation
along the merger axis. The top-right panel shows the kinetic energy
flux-weighted Mach number.  In addition, we masked out all pixels that
contributed $< 10^{-10}$ of the peak radio emission.  This was done to
eliminate some of the shocks along the line of sight that are external to the
cluster.  The bottom-right panel shows a projection of the absolute magnitude
of the magnetic field, weighted by density.  Here we weight by density since in
many regions there is no relic emission,which would lead to
difficult-to-interpret values in the those regions.  Notice that the magnetic
field also peaks in the center of the cluster around $1~\mathrm{\mu G}$. There
appears to be a drop in field strength outside the radio relics.  This makes
sense because the merger shock is just reaching those regions, and then
compresses the field behind the shock.

We now use these four quantities to produce the images in the middle bottom
column.  First, the middle-bottom image shows the radio emission calculated as
outlined in Section \ref{sec:synchrotron_emission}.  Notice that as expected,
the radio emission traces the regions that combine all four of the quantities
in the left and right columns.  This snapshot of the cluster properties suggests
that \textbf{radio emission requires a combination
of dense, hot, magnetically threaded gas in the presence of moderately strong
shocks}.  There are regions in the Mach panel of shocks in the $\Mach=5-7$
range that only contribute a small amount of radio emission.  This region,
which can be seen by rotating the viewpoint (not shown in this presentation),
happens to lie outside the cluster a bit further where the magnetic field,
temperature, and density have dropped considerably.  We also notice that the
strongest regions of radio emission correspond to Mach numbers in the $3-5$
range with temperatures above $10^8 K$, similar to findings in
\citet{2011ApJ...735...96S}. 

Finally, in the top-middle panel, we have overlaid the radio emission on top of
the thermal Sunyaev Zel'dovich effect (tSZE). We have taken out the frequency
dependence in the $f(x)$ function, and multiplied the tSZE Compton y-parameter
by the temperature of the CMB in order to get units of $\mathrm{mK}$. Finally
we multiply by $2.0$, the maximum of $f(x)$, to get the maximum decrement
value.  Notice the very strong correlation with the jump in tSZE and the
presence of radio emission.  Also note that this image is shown in
linear-scale, and the dynamic range in this image is $<20$.  This has
interesting implications for ongoing and future high-resolution tSZE
measurements with MUSTANG \citep{2011MmSAI..82..485M}. CARMA
\citep{2012arXiv1203.2175P}, and CCAT \citep{2009ASPC..417..113R}.  Since the
tSZE is sensitive to the integral of the pressure along the line of sight, it
may be preferable to X-ray studies of galaxy cluster shocks because of it's
linear dependence on density and temperature instead of a roughly quadratic
dependence on density and sub-linear dependence on temperature.  This also
minimizes the effects of gas clumping and increases sensitivity to low density
gas.  This advantage is amplified in the outer regions of galaxy clusters where
many of these radio relic shocks are located.

\subsection{Density \& Magnetic Field Strength Relationship}

In the next two sections we will describe the physical properties of the radio
emitting plasma in order to compare to what has been found observationally.  In
the following analysis, we use the inner-most nested refined region of the
simulation $(32~\mathrm{Mpc/h})^3$ and ignore the lower-resolution regions in
the remainder of the simulation.  We choose to study this entire sub-volume
instead of regions defined by the virial radius because the cluster is
undergoing a major merger. Within this region, we bin various quantities such
as radio emission, density, temperature, and magnetic field.

\begin{figure}[htp] \centering
    \includegraphics[width=0.5\textwidth]{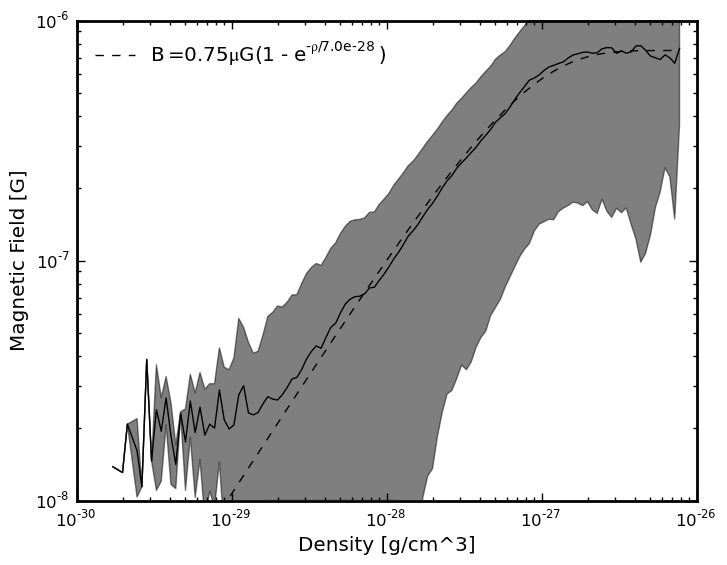} \caption{Average
        magnetic field strength as a function of density, where each cell is
        weighted by the density, with the shaded area denoting the standard
        deviation.  For reference, we show an analytic function that is linear
        with $\rho$ at low density and flattens at high
    density.}\label{fig:mag_v_rho} \end{figure}

Before we discuss the radio emission, we first describe the structure of the
magnetic field.  In Figure \ref{fig:mag_v_rho}, we show the average magnetic
field strength as a function of density, along with the $1-\sigma$ standard
deviation where the average is weighted by density.  This is similar to plots
found in \citet{2008A&A...482L..13D}, though in their study they also followed
gas cooling.  Contrary to what has been
assumed in prior studies such as \citet{2008MNRAS.391.1511H,
2011ApJ...735...96S}, we find that the magnetic field does not scale with
$\rho^{2/3}$, but instead with $\rho$ at low density and has an inner core with
roughly flat magnetic field at high density.  This was discussed previously in
\citet{2011ApJ...739...77X}, casting doubt on assumptions of any tight
relationship between density and magnetic field strength.  This suggests that
the manner in which we inject the magnetic fields may ultimately determine
quantities such as the total radio luminosity.  Future work should explore the
effects of varying the magnetic field injection mechanism.  For example, we may
expect a shallower slope and more evenly distributed field strength if we
inject magnetic fields from multiple sources instead of seeding each cluster
with a single source.

Because of the steep drop in magnetic field strength at low densities, we
expect that the radio emission will similarly decrease towards the outskirts of
the cluster.  Additionally, because the field strength flattens out at high
density, we expect to see radio emission that is not necessarily biased to the
highest density regions near the center of the cluster.

\subsection{Kinetic Energy \& Radio Emission Distributions} 

The next quantity we will use to describe the gas properties of the cluster are
the kinetic energy flux through shocks and the radio emission based on the
\citet{2007MNRAS.375...77H} model.  We find it useful to show both of these
quantities with respect to density, Mach number, magnetic field strength, and
temperature in order to characterize the gas that is most responsible for the
conversion of kinetic energy to cosmic-ray electrons.  For this particular
study, we choose to show both kinetic energy flux as well as the radio
emission.  As seen in Section \ref{sec:synchrotron_emission}, the radio
emission is a complex function of many quantities, and it is difficult to
disentangle each variable.  The kinetic energy flux, however, is a much simpler
quantity, relying only on the density, velocity, and temperature of the
incoming gas: $F_{KE}=0.5 \rho v^2 \Mach c_s$. 

\begin{figure*}[htp] \centering
    \includegraphics[width=1.0\textwidth]{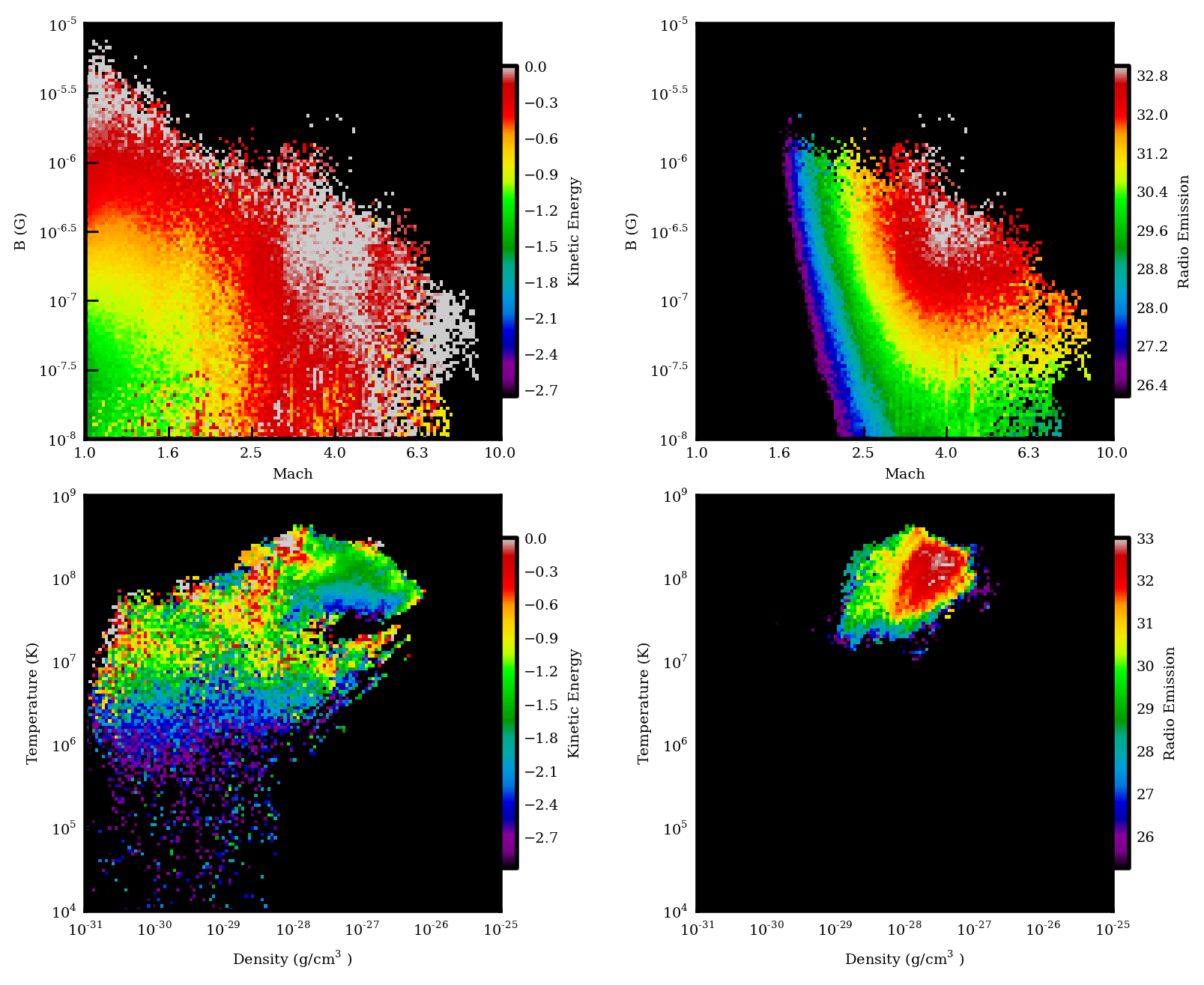} \caption{Phase
        plots of gas properties indicating the location of the kinetic energy
        flux and radio emission at shock fronts.  The left plots show the
        kinetic energy distribution, while the right plots follow the radio
        emissivity.  The top panels show the distributions as a function of
        magnetic field strength on the y-axis, and Mach number on the x-axis.
        The lower panels show them as a function of temperature on the y-axis,
    and density on the x-axis. } \label{fig:phase_4_panel} \end{figure*}

The top left panel of \ref{fig:phase_4_panel} shows the kinetic energy flux
through shocks as a function of Mach number on the x-axis, and the magnetic
field strength on the y-axis.  Note that all un-shocked cells in the volume are
ignored.  For each bin, we calculate the kinetic energy, and normalize by the
maximum value from all the bins.  This distribution is shown in logarithmic
space, as only a small number regions dominate the kinetic energy flux.  In
this figure there appears to be 3 primary populations of gas that contribute to
the kinetic energy flux.  The first is a very low Mach number ($\Mach < 1.5$),
high-magnetic field, region of the simulation, which corresponds to the
turbulent flow in the center of the galaxy cluster.  This slightly supersonic
flow processes a large amount of kinetic energy because of the high density and
high temperature (and therefore sound speed), even though the Mach numbers are
low.  On the upper end of the shock Mach number scale, there are shocks around
a Mach number of 6-8 with a magnetic field strength below $0.01~\mathrm{\mu
G}$.  These, as we will discuss later, are associated with shocks onto
filaments. 

The third population in the kinetic energy flux is at Mach numbers between
$3-5$ with a magnetic field strength of between $0.1-1.0~\mathrm{\mu G}$.
These are the two primary merger shocks, which we will see are the main source
of the radio emission.  Note that this is not a single or even pair of distinct
Mach numbers, meaning that \textbf{observationally, a given merger shock should
not be characterized by a single Mach number}. If the properties of
shock-accelerated electrons are strongly dependent on the Mach number,
ascribing a single Mach number may lead to inconsistencies in the fitting to
the observed radio emission.

The lower-left panel shows the same quantity in color, but now decomposed into
density and temperature bins.  Here we get a different view of the same result.
In this panel there are several knot-like regions in phase space, which likely
correspond to each of the primary shocks in our cluster.  However, in this case
we see that there are 3-4 regions.  There is a very low-density,
$10^6-10^7~\mathrm{K}$ region that likely corresponds to outer accretion shocks
onto the filaments.  The other regions of high kinetic energy at more
intermediate densities correspond to each of the primary merger shocks in the
cluster.  The large clump at very high temperatures and a density of roughly
$10^{-28}~\mathrm{g/cm^3}$ corresponds to the left-moving shock at $\Mach \sim
4$.  This will become even more apparent when we examine the right panels of
this figure.  The primary results of this study of the kinetic energy flux
agrees well with prior studies of the distribution of kinetic energy flux in
shocks \citep{2003ApJ...593..599R, 2006MNRAS.367..113P, 2008ApJ...689.1063S,
2009MNRAS.395.1333V, 2011MNRAS.418..960V}.  However, this is the first time
they have also been correlated with the magnetic field distribution.

Now we contrast the distributions in kinetic energy flux with the radio
emission.  In the right panels of Figure \ref{fig:phase_4_panel}, we use the
color scale to signify the total radio emission.  In the top right panel, we
again show radio emission as a function of Mach number and magnetic field
strength.  Here we see that of the three features present in the kinetic energy
flux, only one remains in the radio emission.  This corresponds to the regions
where the three ingredients needed to create radio emission are all present.
The Mach number is above the threshold for accelerating high energy electrons,
the gas is dense enough, and the magnetic field strength is high enough.  We
note that the falloff at low Mach numbers is due to the functional form of
$\Psi(\Mach)$ from Equation \ref{eq:2}.  For this simulation, we find that
\textbf{the majority of the radio emission in this cluster comes from shocks
with Mach numbers of $3.0-6.0$ and magnetic field strengths of
$0.1-1.0~\mathrm{\mu G}$.}    

The same general conclusions are found from the lower right panel. We see that,
instead of a fairly broad distribution in kinetic energy flux across densities
and temperatures, the radio emission is confined to regions of relatively high
densities and temperatures near $10^8~\mathrm{K}$. By combining this with the
upper-right panel, we have determined the exact makeup of the gas responsible
for the synchrotron radiating electrons in radio relics.

It is important to note that the magnetic field in our simulation does not
reach the levels calculated from observations of several radio relics
\citep{2001ApJ...547L.111C, 2004JKAS...37..337C, 2010Sci...330..347V}.  This suggests that the
magnetic field injection mechanism used in this study may not correspond to how
it occurs in observed galaxy clusters.   Alternatively, amplification of
pre-existing fields prior to AGN injection, primordial magnetic fields, and
several other processes such as turbulent dynamo \citep{2012MNRAS.423.2781I} in
the post-shock region and the streaming instability
\cite[e.g.][]{2007A&A...475....1A} may lead to higher magnetic field values, all
of which could be investigated in future work. Finally, observations should be
re-examined to determine if a lower value of magnetic field is possible, such
as what has recently been found in \citet{ 2012arXiv1205.1082C}.

\subsection{Merger Evolution}

\begin{figure*}[htp] \centering
    \includegraphics[width=1.0\textwidth]{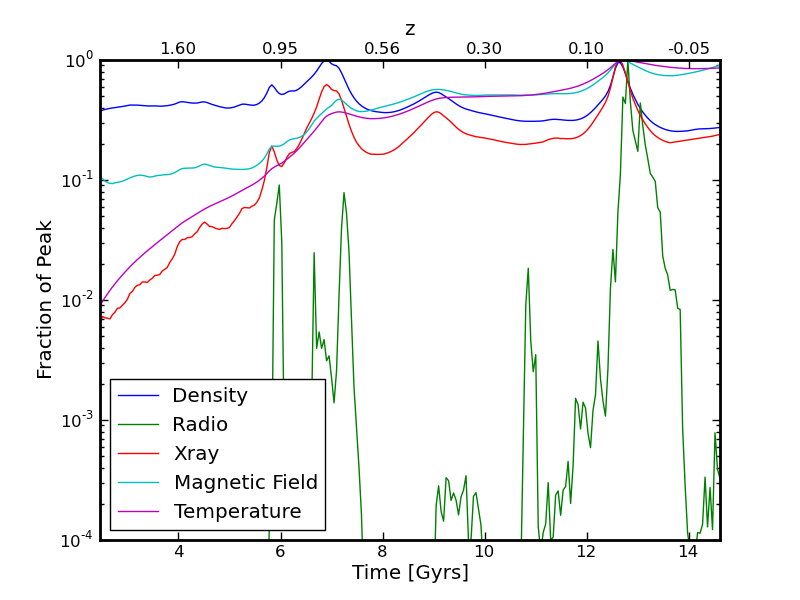} \caption{Time
        evolution of integrated gas properties in the volume surrounding the
        structures of interest.  For density, temperature, and magnetic field,
        we calculate a weighted average, using density as the weight.  For
    radio and X-ray values, we calculate the total emission within the
innermost nested region of the simulation. Note the simulation was run beyond
$z=0$ to allow the merger to finish.}\label{fig:time_evolution} \end{figure*}

Observationally, we are limited to a single snapshot in time, from which we
must deduce the prior and future evolution of a given galaxy cluster.
Fortunately this is not the case for simulations.  In this section we describe
in detail the evolution of the gas properties throughout the history of the
clusters we are modeling.  Figure \ref{fig:time_evolution} shows the evolution
of bulk properties in the nested region of the simulation from redshift
$z=2.95$ to $z=0.0$.  There are five quantities shown here.  \begin{enumerate}
\renewcommand{\labelitemi}{$\bullet$}
  \item Density: Average density, weighted by density.
  \item Temperature: Average temperature, weighted by density.
  \item Magnetic Field: Average magnetic field strength, weighted by density.
  \item Radio: Total 1.4 GHz Radio Emission
  \item Xray: Total 0.5-12 keV X-ray luminosity
\end{enumerate}
In each case we select the inner-nested region of the simulation, and use
WeightedAverageQuantity or TotalQuantity ``derived quantities''.  Each of these
averages and totals are then saved for future analysis.  In the case of Figure
\ref{fig:time_evolution}, we normalize each of the quantities by their maximum
value in order to fit them all on the same scale.  In the case of the X-ray and
radio luminosity, the emission is calculated using the blueshifted frequencies
as these would be redshifted into the observer's frame to be at the correct
values ($0.5-12 \mathrm{keV}$ and $1.4 \mathrm{GHz}$ for X-ray and radio,
respectively).

We find a very interesting correlation with all of the fundamental and
derived quantities.  First, it is clear that there was not only the late-time
merger near $z=0$, but also earlier merger evolution near $z=1$.
First, we see that the density and temperature
both start to rise $0.2-0.5 \mathrm{Gyr}$ before the radio emission spikes.
Analogously, the magnetic field and X-ray luminosity also follow this slow rise
to a peak.  Near the peak, the radio emission jumps up several orders of
magnitude.  This corresponds to the formation of the shock front that then moves
outwards from the cluster center towards the outskirts of the cluster. 
After the core passage, the density and temperature returns to a lower but
elevated level with respect to the pre-merger values.   

This suggests that the radio emission lags the merger event by a few $\times
10^8$ years while the shock is setting up and expanding into the intracluster
medium.  It again highlights the dependence on not only the local
characteristics of the emitting plasma, but also the shock surface area.
Another key point is that the short timescales over which the radio luminosity
varies implies that for a given mass or X-ray luminosity, there may be very
large scatter in the radio luminosity.  Overall, the radio emission is only
above $10\%$ of its peak value for $\sim 0.5~\mathrm{Gyr}$.  This could help
explain the observed lack of radio emission from clusters with obvious mergers
such as Abell 2146 \citep{2011MNRAS.417L...1R}.  One caveat to this result is
that we do not follow the electrons as they cool.  However, because the cooling
timescale, 
\begin{eqnarray} \label{eq:electrons}
        \tau \approx 2\times10^{12} \gamma ^{-1} ((1+z)^4 + (B/3.3\mu
        G)^2)^{-1}~\mathrm{years}
\end{eqnarray} 
for these electrons with $\gamma \sim 1000-5000$ is short \citep{2008SSRv..134...93F}, 
this additional time has little effect.  Therefore the
characteristics of this time evolution should not change substantially with a proper
treatment of the aging electron population.

\section{Observational Implications}

In this section we set out to provide a theoretical perspective on the analysis
of observed radio relics.  In particular, we comment on some of the assumptions
that are often employed, and in several cases point out how these may be
dubious.  We begin by examining how the viewing angle of a radio relic can
impact its interpretation. We expand on this point in the context of spectral
index analysis, where not only does viewpoint play a role, but small-scale
fluctuations in the shock properties can lead to observational signatures that
mimic an aging population of cosmic ray electrons.  Finally, we point out the
limitations of our models; specifically, we must include aging populations of
electrons in future calculations in order to capture accurate polarization
fraction and direction.

\subsection{Viewpoint}\label{sec:viewpoint}

\begin{sidewaysfigure*}[htp] \centering
\vspace{100 mm}
  \includegraphics[width=1.0\textwidth]{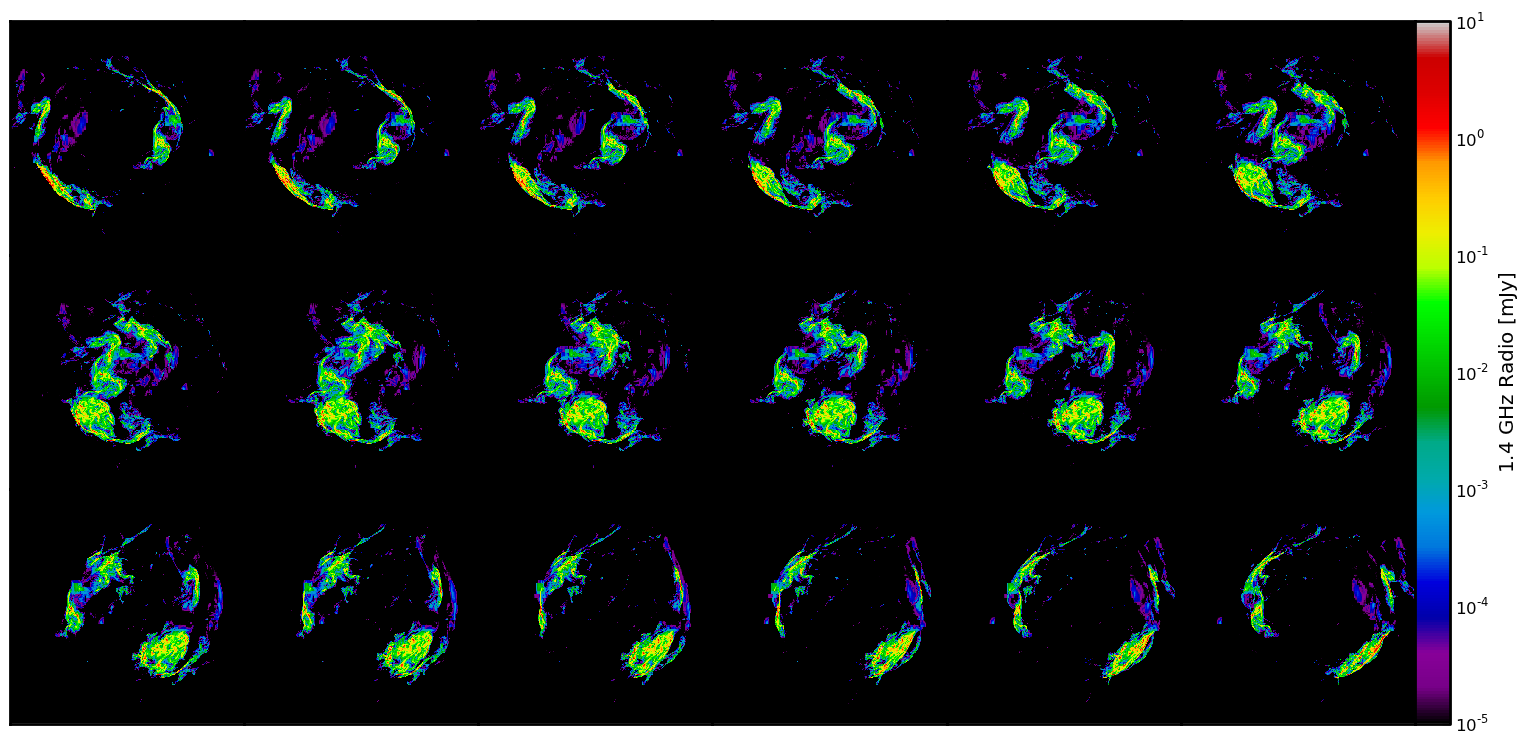}
  \caption{180 degree rotation of radio emission.  Flux computed by assuming a
          distance of 100 Mpc. Each panel is rotated by 10 degrees from the
          previous, moving left to right, top to bottom.  Note that this panel of
    images acts as a autostereogram, and can be used to see the 3D structure of
    the radio emission. For an animation, see
    \url{http://www.youtube.com/watch?v=WNez-h3uOPA}}
  \label{fig:rotation} 
\end{sidewaysfigure*}

Observationally, we are limited to a single viewing angle for each object.
Unfortunately, because radio relic emission is not spherically symmetric, there
will be a viewing angle-dependent emission strength and morphology.  In this
section we set out to demonstrate this fact and how it can affect our
interpretation of radio emitting regions in galaxy clusters.  We begin by
taking our original viewpoint and rotate the viewing angle by 180 degrees over
18 frames.  The result is shown in Figure \ref{fig:rotation}, where radio
emission is projected along each viewing angle. 

What can be seen from Figure \ref{fig:rotation} is that, while for some
orientations the radio emission forms an obvious ``double relic''
configuration, other orientations yield what seems to be a single, more
diffuse, object.  However, given a single frame, it would be difficult to
determine the true structure of the emission.  In fact, this may lead to
a misinterpretation of the radio emission to be that of radio halo origin.  We
will expand upon this line of reasoning in Section \ref{sec:spectral_index}.
There are several other things to note here in the rotation of the radio
emission.  We see that in some orientations (see top row, 5th \& 6th columns),
we reproduce morphologies that include two outer relics, with one of the relics
ending up between the two, close to the center of the cluster.  This is very
similar to observed clusters such as $\mathrm{MACS\ J1752.0+4440}$
\citep{2012MNRAS.425L..36V}, $\mathrm{CIZA J2242.8+5301}$
\citep{2011MmSAI..82..569V}, and  $\mathrm{MACS\ J0717.5+3745}$
\citep{2009A&A...505..991V}.  It may be possible that the emission in these
clusters be not of radio halo origin, but simply radio relic emission viewed
coincident with the cluster center.  Similar work has been done with
hydrodynamic simulations of galaxy clusters and viewing them along the
coordinate axes in \citet{2012MNRAS.421.1868V}, where they found that the low
emissivity due to the small size along the line of sight may explain the lack
of central radio relics.


\begin{figure*}[htp] \centering
    \includegraphics[width=1.0\textwidth]{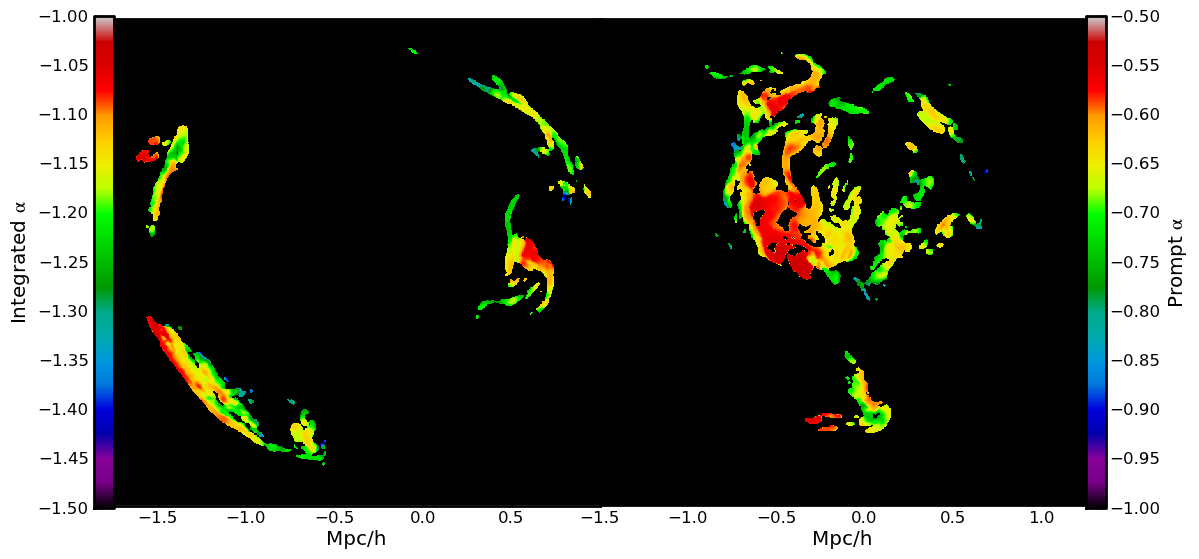} \caption{Spectral
        index of simulated radio relic emission. The left portion of the image
        shows the ``edge-on" view, whereas the right shows the ``face-on" view.
        Both views are on the same scale.  The left colorbar shows the mapping
        of color to the integrated spectral index including particle aging.
        The right colorbar shows the mapping of color to the prompt spectral
        index.  Both colorbars apply to both views, providing a rough estimate
    of the uncertainty in our models of the spectral
index.}\label{fig:compare_alpha} 
\end{figure*}

\begin{figure*}[htp] \centering
  \includegraphics[width=1.0\textwidth]{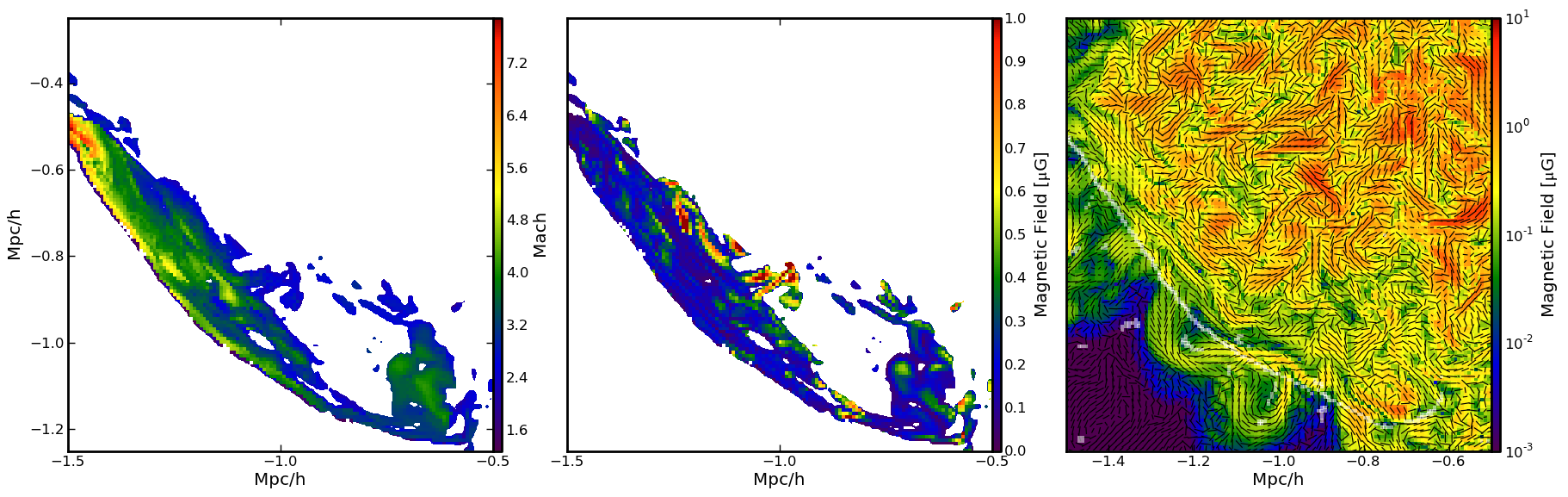} 
  \caption{A zoom in of the lower left radio relic. The left two panels show
      radio emission-weighted projections of the Mach number (left), and magnetic
      field strength (middle). The right panel shows a slice of the magnetic
      field strength, with black lines indicating the local magnetic field 
      direction in the plane of the slice and the white overlay show the 
      location of cells identified to be shocks.}
  \label{fig:bslice} 
\end{figure*}

\subsection{Spectral Index} \label{sec:spectral_index}

Observationally, the spectral index of cluster radio relics is measured by
comparing the emission at several different observed frequencies.  In our work,
we calculate the spectral index by first calculating the radio emission at two
frequencies (here we use 1.4 GHz and 330 MHz).  After smoothing by a Gaussian
with a full-width half maximum size of 4 pixels, we then use these two maps to
calculate the spectral index, similar to what would be done observationally.
Because we use the \citet{2007MNRAS.375...77H} model, we will recover
a spectral index from comparing these two maps that is equal to the cumulative
spectral index due to the emission from electrons over their entire lifetime.
This is steeper than the spectral index that would be predicted from the prompt
emission from a shock front, which is related by $\alpha_{prompt}
= (1-2\alpha_{integrated})/2$.  

In reality, what is thought to happen is the leading edge of the shock front
should accelerate electrons to $\alpha_{prompt}$.  As the radio emitting
electrons move downstream from the shock they cool and the spectrum
steepens. Qualitatively, we would expect that ``edge-on'' observations should
produce values near $\alpha_{prompt}$, whereas a ``face-on'' view would be
sensitive to the entire lifetime of the electrons, and therefore be closer to
$\alpha_{integrated}$. 

Spectral steepening is found to happen in several observed clusters
\citep{2011MNRAS.418..230V} and measuring the steepening of the electrons as
they progress away from the shock can help constrain the local magnetic field,
as was done in \citet{2011MNRAS.418..230V}.  However, the assumption in these
calculations is that the spectral steepening is due entirely to the aging of
the electrons.  However, we see similar steepening even though we do not
include the spectral aging of the electrons as they advect downstream!  These
variations are entirely due to a varying magnetic field and shock strength in
a non-uniform medium.  The spectral indices of the prompt and integrated
spectra are shown in Figure \ref{fig:compare_alpha}, and the fluctuations in
the underlying fields for the lower-left ``edge-on'' relic are shown in the
right panel of Figure \ref{fig:bslice}.

In Figure \ref{fig:compare_alpha}, we see that the ``edge-on'' view shows
a spectral steepening whereas the ``face-on'' view shows the spatially variant
spectral shape due to the shock properties.  For completeness we have mapped
the colormap of both viewing angles to both the prompt and integrated spectral
indices.  The same qualitative behavior is seen with both calculations. These
fluctuations are explained in Figure \ref{fig:bslice}, where we see that the radio
emission-weighted Mach number can vary between $2-8$, and the magnetic field
can vary from $0.1-1 \mu G$.  \textbf{Therefore we warn that extrapolating from
the measured spectral steepening to calculate gas properties should be done
with care, as projection effects and spatial variation of the gas properties
are important.  It may be insufficient to prescribe a single Mach number or
magnetic field to a given radio relic.}  

\subsection{Polarization Fraction \& Direction}

Using our newly developed method, we have created radio polarization fraction
and direction maps for our galaxy cluster from two viewing positions.  The
first gives an ``edge-on`` view of the radio relic, whereas the second is
aligned such that the relics are viewed ``face-on.''  We will examine, in
detail, the polarization signatures of the ``edge-on,'' and then describe the
differences in the ``face-on'' view.

\begin{figure*}[htp] \centering
  \includegraphics[width=1.0\textwidth]{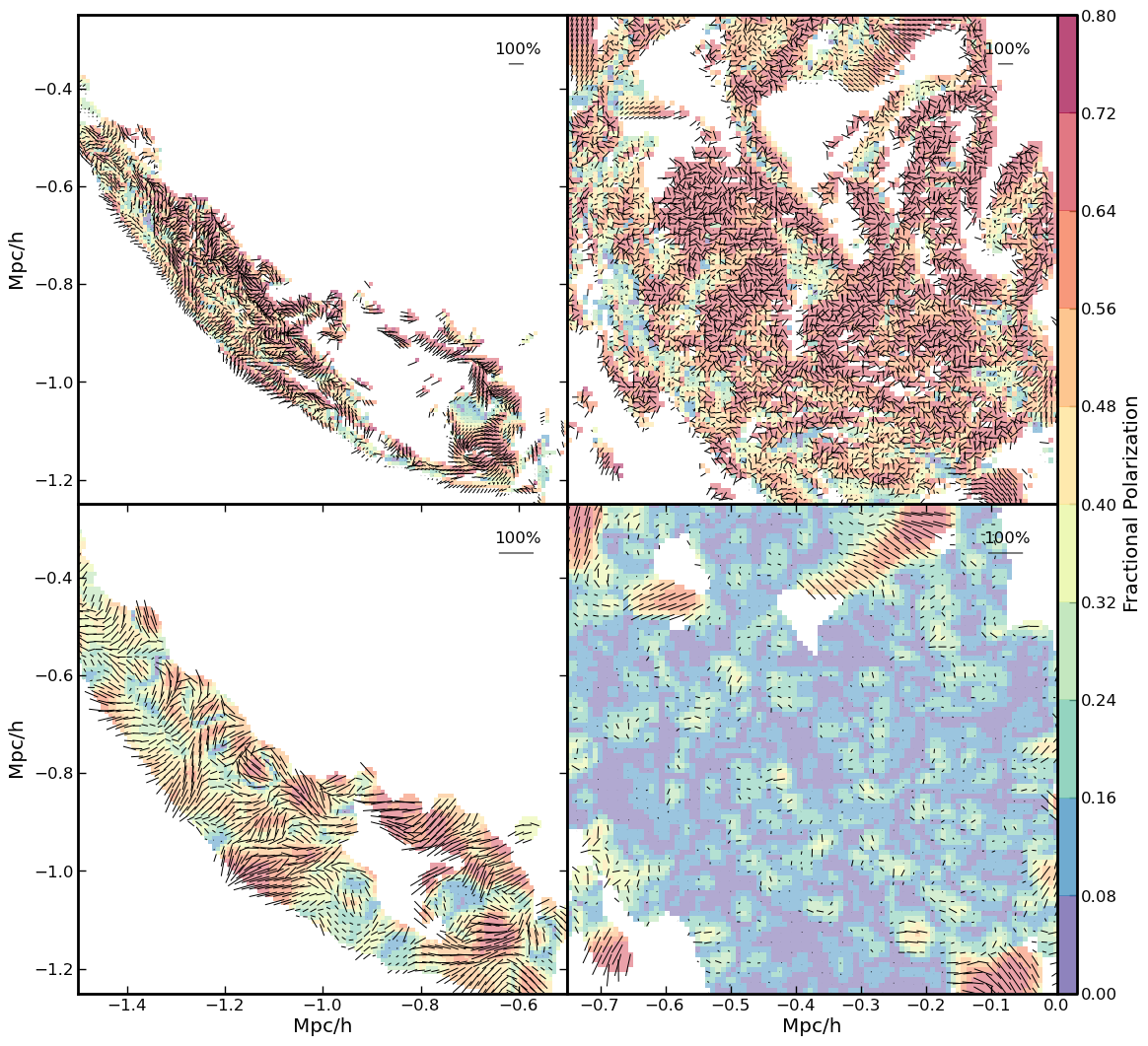}
  \caption{Polarization fraction and direction.  In each panel, the
      polarization direction is denoted by the black quivers, while the
      polarization fraction is represented by both the color scale as well as
      quiver length.  The top panel shows the relic at full resolution $(7.8
      \mathrm{kpc/h})$, while the lower panels show the same view at 4 times
      worse resolution.  At $z=0.2$, the redshift of CIZA J2242.8+5301, this
      corresponds to angular resolutions of $3.36''$ and $13.44''$,
      respectively.  The left panels show the polarization for the ``edge-on"
      (top) while the right shows the ``face-on" view.
  } \label{fig:pol_fraction} \end{figure*}

In Figure \ref{fig:pol_fraction}, we show the linear polarization fraction and
direction of $1.4\mathrm{GHz}$ radio emission.  Lines denote the local linear
polarization direction, the length of which corresponds to the polarization
fraction.  For clarity, the polarization fraction is also shown in color.  The
polarization fraction is calculated using 
\begin{eqnarray} 
    f_p = \frac{\sqrt{I_x^2 + I_y^2}}{I} 
\end{eqnarray}
In this case we
use a $512\times512$ pixel image plane to project through a cube of length
$4~\mathrm{Mpc/h}$ on a side. At full resolution $(7.8 \mathrm{kpc/h})$, we
see that the polarization fraction reaches a maximum of $\sim75\%$, and that
the polarization direction is correlated along the relic, primarily
perpendicular to the shock.  This is very similar to what is found
observationally in CIZA J2242.8+5301 \citep{2011MNRAS.418..230V}.  The authors
find strong (50-75\%) polarization at what is presumed to be the leading edge
of the shock.  Additionally, they find that the polarization direction is
fairly constant over the length of the relic.  We do, however, see greater
variation in the polarization fraction and direction both across and along the
relic.  These fluctuations in our simulation may suggest that there are
additional physical processes which lead to a more ordered field. In order to
investigate the small scale fluctuations in the polarization direction, we
examined a slice of the magnetic field strength and direction through this
relic.  This is shown in Figure \ref{fig:bslice}. What we found is that while
the shock (overlaid in white) cuts through regions which have fairly strong
variations in the magnetic field direction, the region ``behind'' the shock has
a magnetic field that is compressed along the shock propagation direction.
Therefore, it may be possible that if we were to follow the evolution of the
cooling electrons as they moved downstream across the shock, the magnetic field
that they reside in may become more ordered parallel to the shock, leading to
a longer coherence length as more constant polarization direction.
\textbf{Therefore we would expect that in future work when we examine the
emission from an evolving population of cosmic rays, we should see even better
agreement with observations of polarization direction.}

In the top-right panel of Figure \ref{fig:pol_fraction}, we show the same
polarization map, but this time for when the relic is viewed ``face-on''.  We
again see a very high polarization fraction.  However, this time the
polarization direction is significantly less coherent. This is due to the
magnetic field not being modified as strongly in the plane of the shock as it
is perpendicular to the shock.  Therefore the turbulent structure is preserved
in the image plane and the polarization vectors are not preferentially
modified.

However, the behavior of the polarization fraction and direction drastically
changes if we then apply a guassian kernel with a size of 4 pixels.  In the
``edge-on'' view, the polarization fraction and direction is fairly well
preserved.  The fraction only drops to between $30-65\%$ and the direction is
still fairly correlated across the relic.  In contrast, for the ``face-on''
view the polarization has dropped to between $0-15\%$ in most regions.  This is
a classic example of beam depolarization.  Because the polarization direction
is highly disordered, smoothing the image drastically reduces the overall
polarization.

Currently our simulations do not exhibit as much of a constant polarization as
that found in observations of clusters such as CIZA J2242.8+5301, where the
polarization direction is constant over Mpc-scale distances.  There are several
possible explanations. Because we are not tracking the electron distribution as
it cools behind the shock, we may be missing the emission from the more ordered
field line regions behind the shock.  Simulations capable of tracking these
electrons are therefore needed to explore that possibility, and will be
addressed in future work.  If doing so is still incapable of producing
Mpc-scale ordered polarization maps, it may suggest that the injection
mechanism or magnetic field evolution is different than what we have simulated. 

\section{Discussion \& Future Directions}

We have carried out high resolution MHD AMR cosmological simulations using an
accurate shock finding algorithm with a radio emission model for
shock-accelerated electrons to examine the properties of radio relics in galaxy
clusters.  We summarize the physical conditions of the cluster, and several of the
warnings for the interpretations of observed radio features:
\begin{itemize} \renewcommand{\labelitemi}{$\bullet$}
  \item Cosmological initial conditions lead naturally to the formation of
      giant radio relics whose properties are very similar to observed relics.
      This is a natural result of the process of mergers that create galaxy
      clusters, where the magnetized intracluster medium is subject to a series
      of strong, large-area shocks.   
  \item We find that the radio emission in our simulated clusters is strongest
      in plasma that has $0.1-1.0~\mathrm{\mu G}$ magnetic fields with a shock
      Mach number between $3-6$, with densities near
      $10^{-28}~\mathrm{g/cm^3}$ and temperatures of $10^8~\mathrm{K}$. 
  \item If the shock acceleration efficiency for electrons is higher than
      assumed at low Mach numbers, a large reservoir of kinetic energy becomes
      available in hot $(10^7~\mathrm{K})$ gas at both lower and higher
      densities than what is currently producing emission in our simulations.
  \item We find a magnetic field distribution that does not follow the often
      assumed $B\propto \rho^{2/3}$ relation. Instead, it follows $B\propto
      \rho$ at low densities, and flattens out to $\mathrm{sub-\mu G}$ levels
      above densities of $10^{-27}\mathrm{g/cm^3}$.
  \item \textbf{Warning:} Without a model of spectral aging of electrons, we
      still recover a spectral index gradient, indicating that observed
      gradients should not necessarily be interpreted as the spectral aging of
      electrons.  It may simply be due to the projection of a curved shock
      front with varying Mach number.
  \item \textbf{Warning:} From the time evolution of our simulated galaxy
      cluster, it is clear that only a small portion of its lifetime may be
      spent in a regime where it is bright in the radio wavelengths. This may
      explain the apparent lack of observed radio emission in some massive
      galaxy clusters with early or late phase merger.
  \item \textbf{Warning:} The viewing angle of a given galaxy cluster may have
      significant impact on the classification of its radio emission.  Double
      radio relics viewed at some orientations are difficult to differentiate
      from radio halo emission.
  \item \textbf{Warning:} Future simulations of galaxy cluster radio relics
      must follow the temporal evolution of the electron population in order to
      reproduce valid polarization results, as the downstream conditions from
      the shock may include more ordered magnetic fields, leading to different
      polarization maps than when the emission is assumed to come only from the
      shock center.
\end{itemize}

Many of these warnings apply both to observational and theoretical
studies, and the classification and analysis of radio relics in all
contexts must be done with care to avoid confusion with radio halo
emission. There are several advancements that can be made
theoretically.  We are in the process of developing and testing the
numerical framework necessary to follow the cosmic ray electron and
proton populations, using a method similar to
\citet{2001CoPhC.141...17M} and \citet{2005ICRC....3..269J}.  Doing so will
allow us to probe the spectral distribution of these non-thermal
populations in the context of high-resolution cosmological
simulations.  Once this is merged with our ability to produce
synthetic emission and polarization maps, we will be able to more
directly compare to current observations.  Additionally, we do not
explicitly include any magnetic field source terms at the shock
fronts. Exploring local field generation mechanisms such as the Weibel
instability \citep{1959PhRvL...2...83W, 1959PhFl....2..337F} may allow
for alternate magnetic field strength and structure behind the shock
front from what we present here.  Finally, we may need to push to
higher resolution studies at the shock fronts to be able to follow the
shock-amplification of magnetic fields.  Doing so in a cosmological
simulation is currently intractable; however, improvements in
computational speed and parallelization may allow for future studies.

Observationally we are entering a golden age of radio telescopes with the
upgraded Jansky VLA, GMRT, and LOFAR, and are looking forward to the SKA, and
possibly a lunar farside radio telescope\citep{2012arXiv1204.3574B, 
2009astro2010T..50L}.  These
improvements will lead to greater sensitivity and bandwidth, allowing for
multifrequency studies of galaxy cluster environments in unprecedented detail.
However, only through a coordinated effort between simulations and observations
will we be able to fully understand the plasma physics in these cosmic
environments.

\acknowledgements{ The authors would like to thank the referee for in-depth
    comments that led to a much improved paper. S.W.S would like to thank
    Matthias Hoeft and Marcus Br\"{u}ggen for making their radio emission model
    available.  E.J.H.  and
    J.O.B. have been supported in part by grants from the US National Science
    Foundation (AST-0807215, AST-1106437).  S.W.S. has been supported by a DOE
    Computational Science Graduate Fellowship under grant number
    DE-FG02-97ER25308.  B.W.O. has been supported in part by a grants from the
    NASA ATFP program (NNX09AD80G and NNX12AC98G) .  H.X. and H.L. are
    supported by the LDRD and IGPP programs at LANL and by DOE/Office of Fusion
    Energy Science through CMSO. DC gratefully acknowledges support from the
    Advanced Simulation and Computing Program (ASC) and LANL, which is operated
    by LANS, LLC for the NNSA. M.L.N. acknowledges NSF AST-0808184, which
    supported the MHD algorithm development. The computations utilized the
    institutional computing resources at LANL. Computations described in this
    work were performed using the \textit{Enzo} code developed by the
    Laboratory for Computational Astrophysics at the University of California
    in San Diego (http://lca.ucsd.edu) and by a community of developers from
    numerous other institutions.  We thank all the developers of the
    \texttt{yt} analysis toolkit and, in particular, Matthew Turk for
    developing the off-axis projection tool. We have used the “cubehelix” color
    scheme from \citet{2011BASI...39..289G}.The LUNAR Consortium
    (http://lunar.colorado.edu), headquartered at the University of Colorado,
    is funded by the NASA Lunar Science Institute (via cooperative Agreement
    NNA09DB30A), and partially supported this research. 
}

\bibliography{mhd_relic}

\begin{figure}[htp] \centering
    \includegraphics[width=0.45\textwidth]{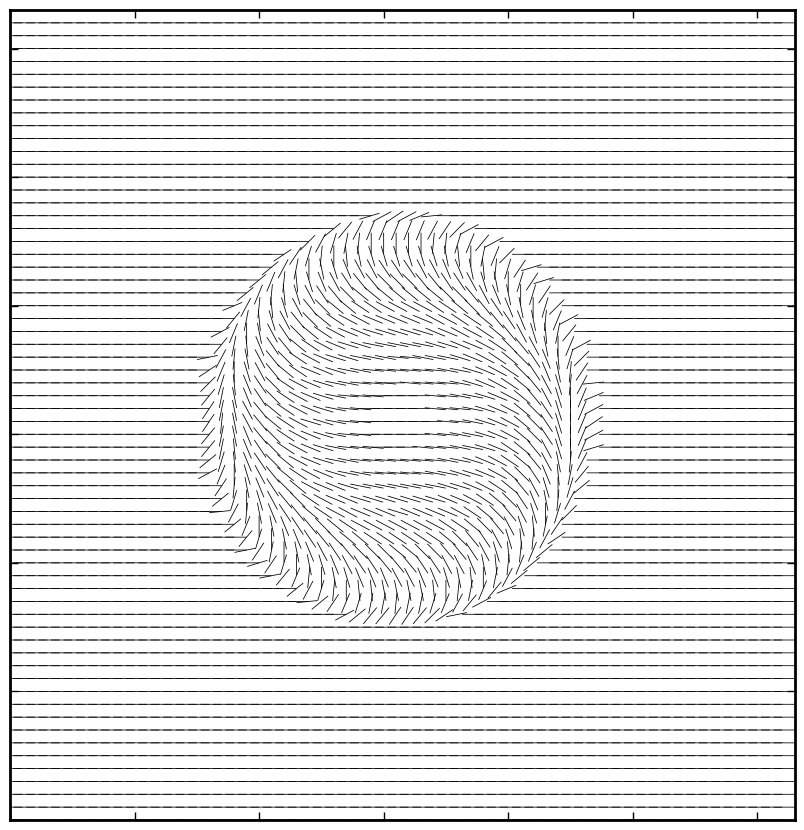}
    \includegraphics[width=0.45\textwidth]{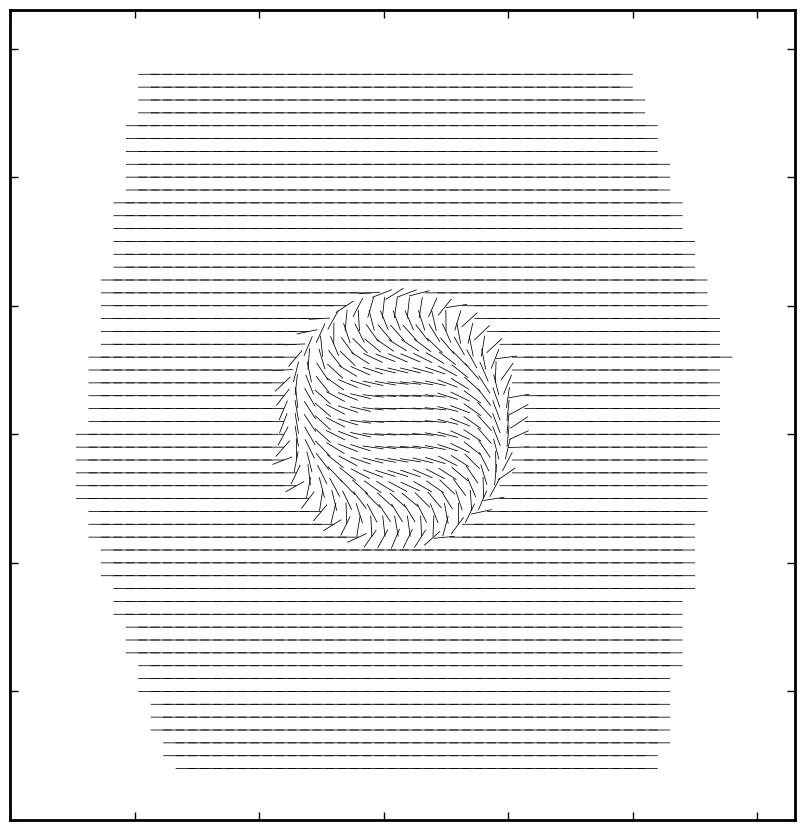}
        \caption{On and off-axis Faraday rotation test.  The left panel
        shows the on-axis Faraday rotation through a magnetized sphere, with an
        image width equal to the domain size.  The right panel shows the same
        rotation, but off-axis and with a width of $1.6$ larger than the left, to
        show the off-axis nature of the domain.  The electron number density
        and magnetic field strength of the sphere are chosen to rotate the
        polarization angle $\pi$ radians for the rays passing through the
        center of the sphere.}
  \label{fig:faraday_test_1} \end{figure}

\begin{figure}[htp] \centering
    \includegraphics[width=0.5\textwidth]{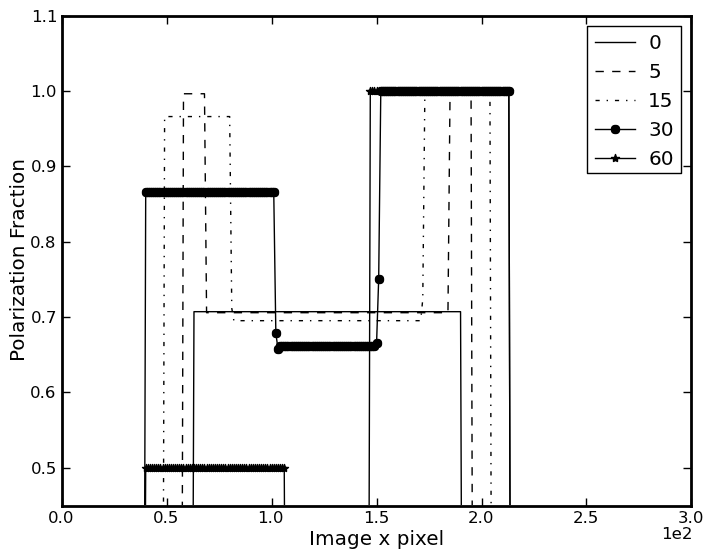}
    \caption{Dual plane polarization test.  The polarization fraction as
    a function of image pixel across the mid-plane of the image, shown for
    varying viewing angles that are measured as an offset in the $x-y$ plane
    from a viewing direction of $\vec{L}=(1, 0, 0)$. }
  \label{fig:dual_plane} 
  \end{figure}

\appendix{
\section{Polarized Emission Integration}

In this section we detail the integration methods used to calculate the total
and polarized radio emission Faraday rotation.  Where possible, we follow the 
conventions listed in \citet{1994hea..book.....L} and \citet{2009ApJ...693....1O}.
As input to the system, we will
assume that there is an emissivity, $\epsilon~[erg/s/cm^3/Hz]$; magnetic field,
$\vec{B}~[G]$; electron number density $n_e~[cm^{-3}]$; and viewing angle,
$\vec{L}$.

From the viewing angle and magnetic field, we first decompose the magnetic
field into components along the line of sight $B_{||}$ and perpendicular to it
$B_{\perp}$.  We use $\theta$ to represent the angle between the $\vec{L}$ and
$\vec{B}$, where we $\theta$ has a range of $0 - \pi$.
\begin{eqnarray} \label{eq:A1}
    B_{||} = \vec{B} \cdot \vec{L} = |B| cos(\theta) \\
    B_{\perp} = |B| sin(\theta)
\end{eqnarray}
We then further decompose $B_{\perp}$ into components aligned with the east and
north vectors in the image plane, here referred to as $B_x$ and $B_y$.  Using
$B_x$ and $B_y$, we then define an angle field, $\chi$,
\begin{eqnarray}
    \chi = (tan^{-1}(B_y/B_x)~+~\pi/4)~\%~\pi
\end{eqnarray}
where we explicitly cast the $tan^{-1}$ into $[0..\pi]$. $\chi$ represents the
local direction of the electric field in the image plane.  We then use this
value to calculate the fractional polarized emission along the east and north
vectors,
\begin{eqnarray}
    f_p = (s + 1)/(s + 7/3) \\ 
    \epsilon_{px} = \epsilon f_p |\frac{B_{\perp}}{B}| cos(\chi)  \\
    \epsilon_{py} = \epsilon f_p |\frac{B_{\perp}}{B}| sin(\chi)  
\end{eqnarray}
where $f_p$ is the polarization fraction, defined using the spectral index of
the relativistic electrons, $s$.

Once these emission terms have been defined, we then integrate through a 
simulation, accounting for the Faraday rotation,
\begin{eqnarray}
    \Delta \psi = 2.62\times10^{-17} n_e \lambda^2 B_{||} dl
\end{eqnarray}
where $\mathrm{dl}$ is in units of cm. All of this is then integrated to
calculate the total intensity $I$, and the polarized components $I_x$ and
$I_y$, which is written using the discrete approximation:
\begin{eqnarray} 
  \left[ { \begin{array}{cc} I_{i+1}  \\ I_{x, i+1} \\ I_{y, i+1}  \\ \end{array} } \right] 
  = \left[ 
    {\begin{array}{ccc} 
        dl &\ 0 &\ 0 \\ 
        dl\ f_p\ (\vec{v_x} \cdot \vec{E}) &\ cos(\Delta \phi) &\ -sin(\Delta \phi) \\ 
        dl\ f_p\ (\vec{v_y} \cdot \vec{E}) &\ sin(\Delta \phi) &\ cos(\Delta \phi)  \\ 
    \end{array} } 
  \right] 
  \left[ {\begin{array}{cc} \epsilon_{i}  \\ I_{x, i} \\ I_{y, i} \\ \end{array} } \right] 
\end{eqnarray} 

where the index $i$ represents the i'th cell along the line of sight.  

As a test of this method, as implemented using \yt, we initialize a completely
polarized background at the simulation domain boundary, with the polarization
angle pointing horizontal in the image plane. To do so, we override the
emission, polarization fraction, density, and magnetic field strengths in the
simulation presented in this paper.  In this case, we set the magnetic field
equal to 0 (though for numerical reasons we use a very small number), except
for a sphere of with a radius equal to one-quarter of the simulation domain
size.  Inside this sphere, we allow the magnetic field parallel to the line of
sight to be exactly that which will, for rays passing through the center of the
sphere towards the ``observer'', be enough to rotate the polarization vector by
$\pi$.  We show the results of this test using $128^2$ pixels in Figure
\ref{fig:faraday_test_1} for both on-axis $(\vec{L}=(1,0,0))$ and off-axis
$(\vec{L}=(1.0,0.3,0.7))$, where we find excellent agreement with the expected
behavior. For both viewing angles, the maximum of $f_p-1.0\approx10^{-14}$, and
the polarization angle has a fractional error in the center of the image of
$1.7\times10^{-5}$ for on-axis, and $1.3\times10^{-2}$ for off-axis. Given the
coarse nature of the sampling of this image and the likelihood that the center
pixel doesn't exactly traverse the center of the sphere, these should be viewed
as upper limits.

As a second test, we initialize two $y-z$ planes of radiation each
with electric field vectors that are perpendicular to each other along. For the
first plane, we set $B_x=B_z = 0.0$, and $B_y=1.0$. The second plane has
$B_x=By=0.0$ and $B_z=1.0$.  Each plane has a width equal to $\frac{1}{32}$ of
the simulation width, centered at $x=\frac{1}{64}$ and $\frac{63}{64}$ in units
of the simulation width.  We set the image width equal to $2.0$ and use $256^2$
pixels.  Finally, we begin by looking from $(-1, 0, 0)$ towards $(1, 0, 0)$,
centered on $(0.5,0.5,0.5)$, and rotate in the $x-y$ plane. Therefore in all
cases plane 2 has a magnetic field that is vertical and plane 1 begins with
a horizontal magnetic field but then proceeds to have a component along the
line of sight.  For this test we disable the Faraday rotation, and examine only
the effects of combining two planes of radiation on off-axis lines of sight.
We vary the rotation from the original viewpoint to have angles of $\theta
= (0, 5, 15, 30, 60)$. In Figure \ref{fig:dual_plane}, we show the fractional
polarization as a function of image pixel. In all cases for the portions of the
planes that overlap along the line of sight, the calculated polarization
fraction is within double precision fractional error of the expected
polarization fraction.

All of these capabilities are demonstrated in a public repository, found using
the changeset with hash fc3acb747162 here:
\url{https://bitbucket.org/samskillman/yt-stokes}.  Future improvements to this
code as well as tighter integration in the primary \yt~repository is expected,
both in performance and usability.  }

\end{document}